\documentstyle[12pt]{article}
\def\mpc{M_{\pi^+}^2}
\def\mpn{M_{\pi^0}^2}
\def\mkc{M_{K^+}^2}
\def\mkn{M_{K^0}^2}
\def\l7{\lambda_7}
\def\la{\langle}
\def\rag{\rangle}

\def\pieta{$\pi^0$-$\eta,\eta'\;$}

\def\epsrat{$\epsilon^\prime/\epsilon\;$}
\def\reepsrat{$\,{\rm Re}\,(\epsilon^\prime/\epsilon)\;$}
\def\optwo{${\cal O}(p^2)\;$}
\def\opfour{${\cal O}(p^4)\;$}
\def\opsix{${\cal O}(p^6)\;$}
\def\ra{\rightarrow}
\def\be{\begin{equation}}
\def\ee{\end{equation}}
\def\bea{\begin{eqnarray}}
\def\eea{\end{eqnarray}}
\def\non{\nonumber}
\def\bb{\bibitem}

\begin{document}
\baselineskip=17pt
\parskip=5pt
  
\begin{titlepage} \footskip=5in     

\title{ 
\begin{flushright} \normalsize 
ISU-HET-99-05 \vspace{0.3em} \\ 
UK/TP 99-09 \vspace{0.3em} \\ 
hep-ph/9909202 \vspace{0.3em} \\ 
September 1999 \vspace{5em} \\  
\end{flushright}
\large\bf   
Additional Isospin-Breaking Effects in ${\bf \epsilon^\prime/\epsilon}$
}  

\author{\normalsize\bf 
S.~Gardner$^{(a)}$\thanks{E-mail: gardner@pa.uky.edu} \  and  
{G.~Valencia$^{(b)}$\thanks{E-mail: valencia@iastate.edu}}  \\
\normalsize\it $(a)$ Department of Physics and Astronomy, 
University of Kentucky, Lexington, KY 40506-0055\\
\normalsize\it $(b)$ Department of Physics and Astronomy, 
Iowa State University, Ames, IA 50011 \\ 
}

\date{}
\maketitle

\begin{abstract}

In the analysis of $\epsilon^\prime/\epsilon$ it has been 
traditional to consider the isospin-breaking effects arising from
electroweak-penguin contributions and from \pieta
mixing, yet additional isospin-violating 
effects exist. 
In particular, we study the 
isospin violation which arises from the $u$-$d$ quark mass
difference in the hadronization of the gluonic penguin operator, 
engendering contributions of an effective $\Delta I=3/2$ character. 
Using chiral perturbation theory and the factorization approximation
for the hadronic matrix elements, we find 
within a specific model for the low-energy constants that we can 
readily accommodate an increase in \epsrat by a factor of two. 

\end{abstract}

\end{titlepage}

\section{Introduction}

The recent measurement of a non-zero value of 
\reepsrat~\cite{ktev} establishes the
existence of direct CP violation in $K\ra \pi\pi$ decays
and provides an important first check of the mechanism of CP violation
in the Standard Model (SM). 
However, the value of \reepsrat which emerges 
from combining the recent KTeV and NA38 results~\cite{ktev}
with  the earlier NA31 and E731 results~\cite{oldeps},
yielding ${\rm Re}\,(\epsilon^\prime/\epsilon)=
(21.2\pm 2.8)\cdot10^{-4}$~\cite{world}, 
exceeds the ``central'' SM prediction of 
$7.0\cdot 10^{-4}$~\cite{burasr,buraslh}
by a factor of three.
This compels us to scrutinize
the SM predictions in further detail: here we study isospin-violating
effects arising from the $u$-$d$ quark mass difference. 

Isospin violation plays an important role 
in the analysis of
$\epsilon^\prime/\epsilon$, for the latter
is predicated by the difference of 
the imaginary to real part ratios in the 
$\Delta I=1/2$ and $\Delta I=3/2$ $K\ra \pi\pi$ amplitudes. 
The differing charges of the $u$ and 
$d$ quarks engender $\Delta I=3/2$ electroweak penguin contributions,
whereas \pieta mixing, driven by the $u$-$d$ quark mass
difference, modifies the relative contribution of the 
$\Delta I=1/2$ and $\Delta I=3/2$ amplitudes in a significant way.
The analyses of Refs.~\cite{dght,buge,extra} consider the effect 
of the electromagnetic penguin operator \cite{bijwi} 
as well as of the mixing of the neutral pion 
with both the $\eta$ \cite{bijwi} 
and the $\eta^\prime$. Recent analyses have
incorporated electroweak penguins in detail, as reviewed in
Ref.~\cite{burasr,bertolini}. 

Here we focus on 
isospin-breaking effects in the gluonic penguin operator.
This operator has always been described as purely 
$\Delta I =1/2$ in nature,
but this is only true in the limit of isospin symmetry. 
That is, although the short distance structure of the 
operator $Q_6$~\cite{tasirafael}, e.g., is manifestly 
$\Delta I=1/2$, 
the differing up and down 
quark masses effectively 
distinguish the interaction of gluons with 
up and down quarks, 
so that the $\langle\pi\pi|Q_6|K\rangle$ matrix element
can possess a $\Delta I=3/2$ component as well~\cite{bpipi}. 
Alternatively, one can consider the $(8_L,1_R)$ operators of
the weak chiral Lagrangian~\cite{cronin,kambor}, which 
embraces operators such as a ``hadronized'' $Q_6$. 
In this case one finds that quark 
mass effects in the octet operators appear at ${\cal O}(p^4)$ in the 
weak chiral Lagrangian; this is an  explicit realization of
the isospin-violating effects we discuss.  

In the isospin-perfect limit, 
\epsrat can be written in terms of the
amplitudes $A_0 \equiv A(K\ra(\pi\pi)_{I=0})$
and $A_2 \equiv A(K\ra(\pi\pi)_{I=2})$
as~\cite{buraslh}
\be
\frac{\epsilon^\prime}{\epsilon} = 
- \frac{\omega}{\sqrt{2}|\epsilon|}\xi(1 - \Omega) \;, 
\ee
where 
\be
\omega \equiv \frac{{\rm Re} A_2}{{\rm Re}\, A_0} \quad;\quad
\xi\equiv \frac{{\rm Im}\, A_0}{ {\rm Re}\, A_0} \quad;\quad
\Omega\equiv 
\frac{{\rm Im}\, A_2}{\omega {\rm Im}\, A_0} \;
\ee
and $\omega\approx 1/22$ emerges 
from an analysis of $K\ra \pi\pi$ branching ratios~\cite{devdic,kmw}. 
The quantity $\xi$ is driven by the gluonic
penguin contribution, and a non-zero $\Omega$ reflects the 
presence of $\Delta I=3/2$ contributions. 

We adopt the notation $\Omega_{\rm IB}$ to denote the contribution
to $\Omega$ generated by the $u$-$d$ quark mass difference\footnote{We 
adopt this notation for simplicity 
and refer the reader to Ref.~\cite{gardner} for a detailed discussion.}, 
which parallels the original discussions of \pieta
mixing effects~\cite{dght,buge}. 
Indeed we have
$\Omega_{\rm IB}= \Omega_{\eta,\eta^\prime} + \Omega_{P}$, where 
the quantity $\Omega_{\rm P}$ 
is driven by isospin violation in the 
hadronization of the gluonic penguin and $\Omega_{\eta,\eta^\prime}$ 
arises from \pieta mixing. Using the isospin decomposition~\cite{dght,buge}
\bea
A({K^0\ra \pi^+\pi^-}) &=& 
 A_0
+ \frac{1}{\sqrt{2}}  A_2 
\label{isodec} \\
A(K^0\ra \pi^0\pi^0) &=& 
 A_0  - \sqrt{2}  A_2 \non \;, 
\eea
and introducing 
``$A_{\rm P}$'' to denote $K\rightarrow \pi\pi$  
amplitudes induced by $(8_L,1_R)$ operators, 
we have 
\begin{equation}
\Omega_{\rm IB} = \biggl({\sqrt{2} \over 3 \omega}\biggr)
\, {{\rm Im} \biggl(A_{\rm P}(K^0 \rightarrow
\pi^+ \pi^-)-A_{\rm P}(K^0 \rightarrow\pi^0 \pi^0)\biggr)\over 
{\rm Im}\,A_{\rm P}(K^0 \rightarrow \pi\pi)} \;
\label{ompdef}
\end{equation}
with ${\rm Im}\,A_{\rm P}(K^0 \rightarrow \pi\pi) = 
({\rm Im}\,A_{\rm P}(K^0 \rightarrow \pi^+\pi^-) + 
{\rm Im}\,A_{\rm P}(K^0 \rightarrow \pi^0\pi^0))/2$. 
The numerator of this expression vanishes 
in the absence of isospin violation. 
Note that a plurality of 
electromagnetic effects, such as final-state Coulomb rescattering
in the $\pi^+\pi^-$ channel~\cite{cirigli}, can also make the right-hand
side of Eq.(\ref{ompdef})
non-zero. We ignore isospin-violating electromagnetic
effects all together, for they are small~\cite{cirigli}, 
 and focus
on the phenomenological consequences of the $u$-$d$ quark mass difference
exclusively. 
At leading-order in chiral perturbation theory, the weak chiral 
Lagrangian does not contain quark-mass-dependent effects~\cite{cronin}
and only $\Omega_{\eta+\eta^\prime}$ is non-zero; however, 
as we show below, the weak chiral Lagrangian does possess such effects 
in ${\cal O }(p^4)$. 

The paper is organized as follows. 
In Section~2 we study the $K \rightarrow \pi\pi$ amplitudes 
at tree level in  ${\cal O}(p^4)$ in the weak 
chiral Lagrangian. This general framework allows us to 
identify all the isospin breaking effects that can occur at 
next-to-leading order in chiral perturbation theory, albeit
the low-energy constants are unknown. 
In Section~3 we consider the gluonic penguin operator. To estimate 
its contributions to the isospin breaking operators we identify 
in Section~2, we use the factorization approximation for the
hadronic matrix elements. Within the factorization approximation, 
the terms of the ${\cal O}(p^6)$ 
strong chiral Lagrangian contribute to the 
${\cal O}(p^4)$ weak chiral Lagrangian. These low-energy 
constants are also unknown; in this case, however, 
we can use a resonance-saturation model~\cite{bijnens} 
to estimate them, to illustrate the effect. 

\section{Chiral Lagrangian Analysis}

Chiral perturbation theory forms a natural framework in which 
to discuss isospin violation in the $K\rightarrow \pi \pi$ amplitudes. 
The weak chiral Lagrangian is realized in terms of the unitary matrix
$U=\exp(2i\phi/f)$, which transforms 
under the chiral group $SU(3)_L\times SU(3)_R$
as $U \rightarrow R U L^\dagger$, where
$R,L$ are elements of $SU(3)_{R,L}$ respectively. 
The function $\phi$ represents the octet of pseudo-Goldstone 
bosons, where $\phi= \sum_{a=1,\dots 8} \lambda_a \phi_a$~\cite{kambor}.
The chiral Lagrangian is constructed in terms of 
$U$ and its derivatives, note that 
$L_\mu \equiv iU^\dagger D_\mu U$, as well as in terms of the function 
$\chi$, which transforms as $U$ under the chiral group. 
In the absence of external fields, $\chi = 2B_0 M$, where $M$ denotes
the quark mass matrix, $M={\rm diag}(m_u,m_d,m_s)$, and the parameter
$B_0$, proportional to the quark condensate $\langle\bar{q}q\rangle$, 
has dimensions of mass. The function $\chi$ encodes the isospin-violating 
effects of interest as $m_u\ne m_d$.

The ${\cal O}(p^2)$, CP-odd\footnote{We thank
G. Colangelo and J. Kambor for their generous assistence 
in rectifying the notational errors of our original manuscript.},
weak chiral Lagrangian transforming as $(8_L,1_R)$ 
under $SU(3)_L\times SU(3)_R$ has only one term~\cite{cronin} 
\begin{equation}
{\cal L}_W^{(2)}=c_2^-\langle\lambda_7 L^2\rangle \;,
\label{lw2}
\end{equation}
where $c_2^-$ is a parameter of order of the Fermi constant $G_F$
and $\langle\rangle$
denotes a trace over flavor indices.  
The $\chi$-dependent term anticipated from the form of the 
leading-order strong chiral Lagrangian, proportional to 
$\la\lambda_7 (\chi^\dagger U + U^\dagger \chi)\rag$, does not
appear in the computation of physical amplitudes~\cite{kambor}. 
The term in question
can be written as a total divergence~\cite{wise} in our case and thus
does not contribute, in accord with
Ref.~\cite{fkw}. 

In next-to-leading order $\chi$-dependent terms
are possible. The complete \opfour weak chiral Lagrangian has been
constructed in Ref.~\cite{kambor}. Collecting the terms which 
can evince isospin violation, we find~\footnote{We choose to 
drop $O_{13}$ as the one of the $O_{10-14}$ 
which is not independent because
it does not contribute to Eq.~(\ref{ompdef}).
Also, $O_{32-34}$ are not included because 
they are related to $O_{1-5}$ by equations of motion.}
\begin{eqnarray}
{\cal L}^{(4)}_{W,{\rm IB}} &=& E_1^- \la\l7 \chi_+^2\rag + 
E_2^- \la\l7 \chi_+\rag\la\chi_+\rag 
+E_3^- \la\l7 \chi_-^2\rag \nonumber \\
&+& E_4^- \la\l7 \chi_-\rag\la\chi_-\rag + E_5^- \la\l7i[\chi_+,\chi_-]\rag
\nonumber \\
&+& E_{10}^- \la\l7\{\chi_+,L^2\}\rag +E_{11}^-\la\l7 L_\mu \chi_+ L^\mu\rag
+E_{12}^-\la\l7L_\mu\rag\la\{L^\mu,\chi_+\}\rag 
\nonumber \\ &+& E_{14}^- \la\l7 L^2\rag\la\chi_+\rag
+E_{15}^- \la\l7i[\chi_-,L^2]\rag \;,
\label{lw4} 
\end{eqnarray}
where $\chi_+$ and $\chi_-$ are defined as 
\begin{eqnarray}
\chi_+ &\equiv& \chi^\dagger U + U^\dagger \chi \nonumber \\
\chi_- &\equiv& i(\chi^\dagger U - U^\dagger \chi) \nonumber \;.
\end{eqnarray}
We can use this Lagrangian to calculate $\Omega_{\rm P}$ 
from Eq.~(\ref{ompdef}). 
Working to leading order in isospin breaking, so that merely terms 
linear in $m_d-m_u$ are retained, and dropping terms suppressed 
by $M_\pi^2/M_K^2$, we find 
\begin{eqnarray}
\Omega_{\rm P} &=& {2\sqrt{2}\over 3\omega}{\mkn\over \mkn-M_\pi^2}
{B_0(m_d-m_u)\over c_2^-}\biggl( 2E_1^- -2E_3^- - 4E_4^- -
E_{10}^- -E_{11}^- -4E_{12}^- -E_{15}^-\biggr)
\nonumber \\
&\approx & {0.12 {\rm GeV}^{2}\over c_2^-}
\biggl( 2E_1^- - 2E_3^- - 4E_4^- -
E_{10}^- -E_{11}^- - 4E_{12}^- -E_{15}^-\biggr) \;.
\label{cptres}
\end{eqnarray}
The terms proportional to $E_1^-$, $E_2^-$, and $E_5^-$ 
can potentially
generate tadpole contributions, as the eigenstates of
the weak interaction are not those of mass. 
Consequently, we take care to remove possible tadpole 
contributions via the 
construction of Ref.~\cite{kambor}. 
Note that to ${\cal O}(m_d-m_u)$ and to the order in the momentum
expansion to which we work, it suffices to use 
Eq.~(\ref{lw2}) to compute ${\rm Im}\,A_{\rm P}(K^0\ra\pi\pi)$. Thus merely
$c_2^-$ appears in the denominator of Eq.~(\ref{cptres}).
The numerical value given reflects the use of 
Ref.~\cite{pdg98} and of the relation
\begin{equation}
B_0 (m_d -m_u) = \mkn-\mkc+\mpc-\mpn
\end{equation}
which follows from the leading-order strong chiral Lagrangian, 
in concert with Dashen's theorem~\cite{dashen,gl}.
The utility of Eq.~(\ref{cptres}) is limited, for the $E_i^-$
coefficients are unknown. However, 
power counting in chiral perturbation theory suggests that each of the 
constants $E_i^-$ is suppressed by 
${\cal O}(\Lambda_{\chi {\rm SB}}^2)$ with respect to $c_2^-$. 
Thus the numerical prefactor in the last line of Eq.~(\ref{cptres}) 
ought determine the ``natural'' size of $\Omega_{\rm P}$ --- 
it is of order $0.1$. 
Remarkably, these effects are of 
comparable numerical size to the value of $\Omega_{\eta}$ in
\optwo~\cite{dght,buge}, so that the terms 
found in Eq.~(\ref{cptres}) merit further study. 

For reference, it is useful 
to summarize the results of Refs.~\cite{dght,buge} for 
$\Omega_{\eta+\eta^\prime}$ in \optwo and then to proceed
to enumerate {\it all} possible isospin-violating contributions in
\opfour, irrespective of whether we term them 
``$\Omega_{\eta+\eta^\prime}$'' or ``$\Omega_{\rm P}$''.
In leading-order chiral perturbation
theory, $\pi^0-\eta$ mixing is the only $m_d\ne m_u$ 
effect to impact the $K\rightarrow \pi \pi$ amplitudes.
Diagonalizing the neutral, non-strange meson states of
the strong chiral Lagrangian in \optwo, noting 
\begin{equation}
{\cal L}_{\rm S}^{(2)} = {f_\pi^2\over 4} \biggl( \la L^\mu L_\mu\rag
+ \la\chi_+\rag \biggr) \;,
\label{ls2}
\end{equation}
yields the physical 
$\pi^0$ state in terms of the octet fields $\pi^0$ and $\eta$:
\begin{equation}
\left(\pi^0\right)_{\rm phys}
= \pi^0 + {\sqrt{3}\over 4}\left({m_d-m_u\over m_s -\hat{m}}\right)\eta 
+ {\cal O}(m_d - m_u)^2\;,
\end{equation}
where $\hat{m}=(m_d+m_u)/2$.
Using Eq.~(\ref{lw2}) to compute
${\rm Im}\,A_{\rm P}(K\ra \pi^0 \eta)/{\rm Im}\,A_{\rm P}(K\ra \pi^0 \pi^0)$ 
yields finally~\cite{dght,buge}
\begin{equation}
\Omega_\eta = {1\over 3\sqrt{2} \omega}\left({m_d-m_u\over m_s - \hat{m}}\right)
\approx 0.13
\label{ometa}
\end{equation}
noting $(m_s - \hat{m})/(m_d-m_u) = 40.8 \pm 3.2$~\cite{leut96}.
The approximate equality of this result to the numerical 
coefficient of Eq.~(\ref{cptres}) follows as 
$4(M_K^2 - M_{\pi}^2) \approx {\cal O}(1\,{\rm GeV}^2)$.
Consequently, it is also important to evaluate the impact of
$\pi^0$-$\eta$ mixing 
in \opfour on the $K\ra\pi\pi$ amplitudes.
This has already been done to some extent, for the usual 
analysis~\cite{dght,buge}
deviates from strict chiral perturbation theory in that 
an explicit $\eta^\prime$ degree of freedom appears as well, 
leading to both $\pi^0-\eta^\prime$ mixing and 
$\eta-\eta^\prime$ mixing. 
For comparison, explicit study of $\pi^0$-$\eta$ mixing in \opfour, 
noting~\cite{gl}
\begin{equation}
{\cal L}_{\rm S,\, IB}^{(4)} = L_4 \la L^2\rag\la\chi_+\rag + 
L_5 \la L^2 \chi_+\rag +
L_6 \la\chi_+\rag^2 - L_7 \la\chi_-\rag^2 + 
{1\over 2}L_8 \la\chi_+^2-\chi_-^2\rag\;,
\label{ls4}
\end{equation}
shows that it is sensitive to the low-energy constant
$L_7$~\cite{gl}. 
In an effective 
Lagrangian which includes the $\eta^\prime$ degree of freedom via 
the nonet symmetry of a large $N_c$ approach, 
taking the limit of small momenta $p^2$ and 
$M_{\eta}^2 \ll M_{\eta^\prime}^2$ 
yields an interaction of
the form associated with $L_7$~\cite{gl}. Moreover, 
the $\eta'$ contribution numerically
saturates the value of $L_7$ found phenomenologically~\cite{gl,reson}. 
The presence of the $\eta^\prime$ thus apes higher-order effects in
the strong chiral Lagrangian. 
Including the $\eta^\prime$ as per the usual 
analysis~\cite{dght,buge} yields\footnote{Note that using the
\pieta mixing formulas resulting from the exact diagonalization 
of a chiral Lagrangian based on nonet symmetry 
in \optwo and large $N_c$~\cite{leuteta} yields 
$\Omega_{\eta+\eta^\prime} = 1.7\, \Omega_\eta \,\approx \,0.22$.
For comparison, note $\Omega_{\eta+\eta^\prime} = 0.25 \pm 0.02$ 
from the recent analysis of Ref.~\cite{hycheng}.
}
\begin{eqnarray}
\Omega_{\eta+\eta^\prime}& = &\Omega_\eta \biggl(
(\cos\theta-\sqrt{2}\sin\theta)^2 +
{M_\eta^2-M_\pi^2\over M_{\eta^\prime}^2-M_\pi^2}
(\sin\theta+\sqrt{2}\cos\theta)^2\biggr)\nonumber \\
&\approx & 2.4\; \Omega_\eta \;\;\approx \;0.31
\end{eqnarray}
where we use $\theta=-22^\circ$ for the 
$\eta-\eta^\prime$ mixing angle~\cite{buge,gl}.  
The effect of the $\eta^\prime$ is no smaller than that 
of the $\eta$; this is consistent with the comparison of Eq.~(\ref{cptres})
with Eq.~(\ref{ometa}).\footnote{Large $N_c$ arguments suggest that
$L_7$ could dominate the low-energy constants
in \opfour \cite{gl}, yet this is phenomenologically not the
case~\cite{gl,reson}.}
There are thus a plurality of effects
which are important in \opfour. Let us enumerate the
possible isospin-violating effects which occur in 
${\cal O}(m_d - m_u)$ and \opfour:
\begin{itemize}
 
\item[i)] 
Isospin breaking in the ${\cal O}(p^2)$ mass term of Eq.~(\ref{ls2}),
including $\pi^0$-$\eta$ mixing, acting 
in concert with the \optwo weak chiral Lagrangian, Eq.~(\ref{lw2}), 
computed to one-loop order. 

\item[ii)] 
$\pi^0$-$\eta$ mixing, 
realized from the ${\cal O}(p^2)$ mass term of 
Eq.~(\ref{ls2}), 
combined with the isospin-conserving 
vertices of the \opfour weak chiral Lagrangian.

\item[iii)] Next-to-leading order $\pi^0$-$\eta$ mixing as per the
strong chiral Lagrangian in ${\cal O}(p^4)$, Eq.~(\ref{ls4}), 
combined with the leading-order weak vertex 
from Eq.~(\ref{lw2}). The $\pi^0$-$\eta^\prime$ 
mixing effects of the usual analysis are an example of this type.
 
\item[iv)] Isospin violation in the vertices of 
the ${\cal O}(p^4)$ weak 
chiral Lagrangian, Eq.~(\ref{lw4}). This is realized as 
Eq.~(\ref{cptres}) and serves as our focus here, for 
it contains the qualitatively new effects we argue.

\end{itemize}

We wish to focus on the contribution of {\it iv)}, yet we cannot 
avoid considering that of {\it ii)},
for the low-energy constants $E_i$ of  
Eq.~(\ref{cptres}) potentially enter here as well. 
Considering {\it exclusively} the terms of Eq.(\ref{lw4}) 
we find the contribution of {\it ii)} to be:
\begin{eqnarray}
\Omega_{\eta + \eta^\prime}^{(4)} &=& 
{2\sqrt{2}\over 3\omega}{\mkn\over \mkn-M_\pi^2}
{B_0(m_d-m_u)\over c_2^-}\biggl(- 2(E_3^- +E_4^- -E_5^-)
+ {1\over 2}(E_{10}^- -E_{11}^-) -2E_{12}^- +E_{14}^-
+\frac{3}{2}E_{15}^-\biggr)
\nonumber \\
&\approx & {0.12 {\rm GeV}^{2}\over c_2^-}
\biggl(- 2(E_3^- +E_4^- -E_5^-)
+ {1\over 2}(E_{10}^- -E_{11}^-)
-2E_{12}^- +E_{14}^- +\frac{3}{2}E_{15}^-\biggr) \;,
\label{cptresad}
\end{eqnarray}
so that there is no manifest cancellation with the terms of
Eq.~(\ref{cptres}). Moreover, we find in Section 3 that the 
contributions of Eq.~(\ref{cptresad}) are numerically smaller
than those of Eq.~(\ref{cptres}) --- it is the latter which 
contains 
isospin-breaking effects in the hadronization of the gluonic penguin 
operator. To study these effects in detail, we 
must turn to the factorization approximation and
estimate, as in the next section, 
the contributions of the gluonic penguin to the $E_i^-$ of Eq.~(\ref{cptres}).

\section{Factorization}

Within the context of the factorization approximation,
the bosonized form of the $Q_6$ penguin operator appears 
as the product of scalar and pseudoscalar 
densities obtained from the strong chiral Lagrangian.
The construction relevant to $K^0 \ra\pi\pi$ decay is~\cite{sekhar,tasirafael}
\begin{eqnarray}
{\cal L}_{\rm P} &=& -{G_F\over \sqrt{2}}V_{us}^\star V_{ud} \, C_6 \,
\biggl( -8 (\bar{s}_Lq_R)(\bar{q}_R d_L)\biggr) ~+~{\rm h.c.} \nonumber \\
&\rightarrow &
 {G_F\over \sqrt{2}}V_{us}^\star V_{ud}\, C_6 \,
32 B_0^2 {\delta {\cal L}\over \delta \chi_{3i}^\dagger}
{\delta {\cal L}\over \delta \chi_{i2}}~+~{\rm h.c.}  \;,
\label{defbos}
\end{eqnarray}
where 
$q_{({L}\,,{R})}=(1 \mp \gamma_5)q/2$ and $C_6$ is defined
as in Ref.~\cite{buclassic}.
Using the \opfour strong chiral Lagrangian~\cite{gl}, 
Eq.~(\ref{ls4}), one finds
\begin{equation}
c_2^- = {G_F\over \sqrt{2}}V^\star_{us}V_{ud}{\rm Im}\, C_6\,\biggl(
16B_0^2 f_\pi^2 L_5\biggr) \;.
\label{fac2}
\end{equation}
Equation (\ref{defbos}) also yields a 
term of the form $\la\l7(\chi_+)\rag$, proportional to $L_8$, 
yet this is merely the weak mass term discussed earlier --- it 
does not contribute here~\cite{wise,fkw}. 
It is well-known that the contribution of the CP-even analogue of 
Eq.~(\ref{fac2}) does not suffice to reproduce the 
phenomenological value of $c_2$, its 
associated low-energy constant~\cite{sekhar}, where we note 
${\rm Im}\, C_6 \ra {\rm Re}\, C_6$ in 
Eq.~(\ref{fac2}) yields $c_2^- \ra c_2$. 
Equation (\ref{fac2}) is useful nevertheless, for 
it serves to normalize the isospin-violating 
constants induced by the $Q_6$ operator.

The \opfour strong Lagrangian, Eq.~(\ref{ls4}), as per
Eq.~(\ref{defbos}), also yields
contributions to certain of the $E_i^-$ coefficients enumerated
in Eq.~(\ref{lw4}), 
as well as to other operators of the \opfour
weak chiral Lagrangian. The non-zero contributions to $E_i^-$ are 
\begin{eqnarray}
E_1^- &=& E_3^- ~=~ -E_5^- ~=~ {2L_8^2 c_2^-\over f_{\pi}^2L_5} \;\quad;\quad
E_2^-~=~  {8L_6 L_8 c_2^- \over f_{\pi}^2L_5} \;\quad;\quad 
E_4^-= {8L_7 L_8 c_2^- \over f_{\pi}^2L_5} \non \\
E_{10}^- &=&E_{15}^-~=~{2L_8 c_2^-\over f_{\pi}^2} \;\quad;\quad 
E_{13}^-= {4L_4 L_8 c_2^- \over f_{\pi}^2L_5} \;\quad;\quad 
E_{14}^-~=~{8 L_6 c_2^- \over f_{\pi}^2} \;.
\label{incfac}
\end{eqnarray}
Unfortunately, however, this approach does not yield a 
complete estimate of the coefficients of the 
\opfour weak chiral Lagrangian in the factorization approximation.
The bosonization of $Q_6$, 
as defined in Eq.~(\ref{defbos}), also contributes to the \opfour weak 
chiral Lagrangian through the ${\cal O}(p^6)$ strong chiral Lagrangian. 
Although the latter has been constructed~\cite{fearing,bijnens}, 
its coefficients are not known, and we must turn to a model to proceed. 

The use of resonance saturation 
allows us to estimate 
some of the coefficients 
in the \opsix  strong chiral Lagrangian.
This has been done for the $L_i$ constants that appear in the 
\opfour strong Lagrangian \cite{reson}. In particular, the 
form of the terms needed in Eq.~(\ref{lw4}) suggest 
that scalar and pseudo-scalar resonances might be dominant. This is 
true for the coefficients of the \opfour strong Lagrangian that 
appear in Eq.~(\ref{incfac}). The constant $L_7$ is saturated by 
the $\eta^\prime$~\cite{gl}, and 
it is reasonable to assume, in an analogous manner, that the 
constants $L_5$ and $L_8$ are saturated
by the scalar resonances~\cite{reson}. 
Indeed, Ref.~\cite{reson} inverts this argument and uses the 
phenomenological values of $L_5$ and $L_8$ to fix the couplings of 
the scalar resonances to the pseudoscalar octet of $\pi$'s and $\eta$'s.

As a model for the needed \opsix counterterms, we 
propose the Lagrangian
\begin{equation}
{\cal L}_S = {1\over 2}\la D^\mu S D_\mu S - M_S^2 S^2\rag
+ c_d \la \xi^\dagger S \xi L^2 \rag + c_m\la\xi^\dagger  S \xi \chi_+ \rag
+{d_m \over 2}\la \xi^\dagger  S^2 \xi \chi_+ \rag 
\label{scalag}
\end{equation}
for the scalar meson octet, 
where $U=\xi^2$.
The first three terms of this Lagrangian are explicitly considered 
in Ref.~\cite{reson}. 
In the limit of momenta such that
$p^2 \ll M_{S}^2$, the scalar octet no longer plays a dynamical
role and is thus ``integrated out,''
yielding \cite{reson}
\be
L_5 ~~=~~ {c_d c_m \over M_S^2} \;\quad;\quad
L_8 ~~=~~ {c_m^2 \over 2 M_S^2} \;.
\ee
Using a scalar mass of $M_S=0.983$~GeV and assuming that 
this contribution saturates $L^r_{5,8}(M_\rho)$, one finds 
$c_m = 0.042$~GeV, $c_d = 0.032$~GeV \cite{reson}.

The last term in Eq.~(\ref{scalag}) has been recently considered 
in Ref.~\cite{bijnens}. This term breaks the mass degeneracy of the states
in the scalar octet, splitting the $K_0^\ast(1430)$ from the 
$a_0(980)$, for example. This term also generates some of the \opsix strong 
operators of interest to us. 
Integrating out the scalar 
octet one finds 
two terms proportional to $d_m$ in \opsix~\cite{bijnens}
\begin{equation}
{\cal L}_S^{(6)}= {d_m c_m^2\over 2 M_S^4}\la\chi_+^3\rag
+{c_d c_m d_m \over M_S^4}\la\chi_+^2 L^2\rag\;,
\label{scalarre}
\end{equation}
which contribute to the scalar densities in the 
bosonization of $Q_6$. The new contributions are
\be
{E_1^-\over c_2^-} ~=~{3 d_m c_m^2 \over 2 M_S^4 L_5} 
\;\; \approx \;\; -4.8~{\rm GeV}^{-2} \;\quad;\quad
{E_{10}^-\over c_2^-} ~=~ {c_d c_m d_m \over  M_S^4 L_5} 
\;\; \approx \;\; -2.4~{\rm GeV}^{-2} \;.
\label{bososca}
\ee
Within the context of our model, other contributions of
the \opsix strong Lagrangian to the constants $E_i$ are assumed to be
identically zero. For the numerical estimates we fit $d_m$ to the 
$K_0^\ast$-$a_0$ mass difference,
\begin{equation}
d_m ={M^2_{K_0^\ast}-M^2_{a_0}\over 2(M_\pi^2-\mkn)} \sim -2.4 \;,
\label{dmval}
\end{equation}
and use $L_5^r(\mu=M_\rho)$ of Ref.~\cite{reson}. 
Note that the terms of Eq.~(\ref{bososca})
are numerically much larger than those of Eq.~(\ref{incfac}) --- indeed,
they dominate $\Omega_{\rm P}$. If we use the values 
of $L^r_{5,7,8}(\mu=M_\rho)$ from Ref.~\cite{reson}: 
$L_5 = 1.4 \times 10^{-3}$, $L_7 = -0.4 \times 10^{-3}$, and 
$L_8 = 0.9 \times 10^{-3}$, Eq.~(\ref{incfac}) yields 
\begin{equation}
{E_1^- \over c_2^-} = 0.13~ {\rm GeV}^{-2}\;;\;
{E_4^- \over c_2^-} = -0.24~ {\rm GeV}^{-2}\;;\;
{E_{10}^- \over c_2^-} = 0.21~ {\rm GeV}^{-2}\;,
\label{trunnum}
\end{equation}
where $f_{\pi}=93\,{\rm MeV}$.
In view of the dominance of the terms computed from the \opsix
coefficients, it is important to compare the relative size 
of the \opfour and \opsix coefficients induced by the scalar  
resonance. To illustrate, let us consider the ratio of the 
coefficient of the first term of Eq.~(\ref{scalarre}), 
calling it $\beta_1$, to the coefficient $L_8$:
\begin{equation}
{\beta_1\over L_8}= {d_m \over M_S^2} \sim -2.4~{\rm GeV}^{-2} \;.
\end{equation}
This ratio is large, but not inconsistent with dimensional analysis.
Nevertheless, it may be naive to associate the 
$K_0^\ast$-$a_0$ mass difference with flavor-symmetry 
breaking as in Eq.~(\ref{scalag}). 
For example, quark model studies suggest that the $a_0(980)$ 
may well be a $K{\bar K}$ molecule~\cite{weinstein,godfrey}.
If we use the predicted lowest-lying isovector and strange scalar states 
of Ref.~\cite{godfrey}, yielding masses of 1.09 and 1.24 GeV,
respectively, we find, rather, that $d_m\sim -0.76$ and
that ${\beta_1/ L_8} \sim -0.79~{\rm GeV}^{-2}$ --- this is
also consistent with dimensional analysis. 

We can now proceed to estimate the value of $\Omega_{\rm P}$
using our estimated low-energy constants. Were we merely to
use the numbers of Eq.~(\ref{bososca}) and $d_m=-2.4$
we would obtain
\begin{equation}
\Omega_{\rm P} = \left(0.12\,{\rm GeV}^2\right)
{d_m c_m(3c_m - c_d) \over M_S^4 L_5} \sim -0.85 \;.
\end{equation}
This unexpectedly large result is driven by 
the value of $d_m$ found in Eq.~(\ref{dmval}): using
$d_m\sim -0.76$ yields $\Omega_{\rm P} \sim -0.28$.
The sign of $d_m$ and thus of $\Omega_{\rm P}$ 
in our picture is the consequence of
the mass of lowest-lying strange scalar being greater than 
that of the lowest-lying isovector scalar. 
Collecting the contributions of Eq.~(\ref{bososca}) and 
Eq.~(\ref{trunnum}) yields 
\be
\Omega_{\rm P} = -0.79 \,(-0.21) 
\ee
for $d_m=-2.4\,(-0.76)$. Note that 
Eq.~(\ref{cptresad}) and 
$L^r_{6}(\mu=M_\rho)=0.2\cdot 10^{-3}$~\cite{reson} yields 
$\Omega_{\eta + \eta^\prime}^{(4)} 
= -0.12\, (-0.03)$, so that writing 
$\Omega_{\rm IB}^{(4)}=\Omega_{\rm P} + \Omega_{\eta + \eta^\prime}^{(4)}$
yields $\Omega_{\rm IB}^{(4)} = -0.91 \,(-0.24)$. 
For reference, the value of Eq.~(\ref{ompdef}) used in the
``central value'' of \epsrat in Ref.~\cite{burasr} is
$\Omega_{\eta+\eta^\prime}=0.25\pm 0.05$, whereas that used
in Ref.~\cite{bertolini} is 
$\Omega_{\eta+\eta^\prime}=0.25\pm 0.10$.
The changes in 
$\Omega_{\rm IB}$ found in \opfour impact \epsrat in a
significant manner. Using 
the simple formula of 
Eq.~(1.7) in Ref.~\cite{burasr} shows that under 
$\Omega_{\rm IB} = 0.25 \ra -0.25$ the value of \epsrat increases by
a factor of $2.2$\,.
Thus a very small or negative value of $\Omega_{\rm IB}$ generates an
increase in \epsrat with respect to the usual value
cited~\cite{burasr}. 
It is particularly noteworthy that the range in our 
estimates of $\Omega_{\rm IB}^{(4)}$ {\it exceed} the 
central value of $\Omega_{\eta+\eta^\prime}$ used in 
Refs.~\cite{burasr,bertolini}. 
The detailed numerical results we find do rely on 
a simple model; nevertheless, 
a substantial increase in the error associated with $\Omega_{\rm IB}$, 
Eq.~(\ref{ompdef}), is in order. 

\section{Conclusions}

We have shown that there are isospin-breaking effects in
$\epsilon^\prime$ which have not been previously 
considered. Specifically, we have examined 
the role of isospin violation in the 
matrix elements of the gluonic penguin operator within the
context of chiral 
perturbation theory. 
Although the presence of unknown low-energy constants implies that
we lack a reliable way to calculate these effects, 
we believe such limitations
underscore the need for a 
larger uncertainty in the theoretical value 
of \epsrat than currently in vogue.
In particular, the recent reviews of 
Refs.~\cite{burasr,bertolini} use $0.25\pm 0.05$ and $0.25\pm 0.10$,
respectively, for the value of 
$\Omega_{\rm IB}=\Omega_{\eta + \eta^\prime}$. 
Our estimate of $\Omega_{\rm IB}$ from 
the specific $m_d\ne m_u$ effects we consider
ranges from $0.1 \ra -0.7$. This range reflects a variation in 
\epsrat of more than a factor of two. 

\noindent {\bf Acknowledgments} The work of S.G. and G.V. is supported 
in part by the DOE under contract numbers DE-FG02-96ER40989
and DE-FG02-92ER40730, respectively. We thank 
J.~F.~Donoghue, C.-J. David Lin, and M.~B. Wise 
for useful conversations, and we thank J.~Bijnens for 
pointing out Ref.~\cite{bijnens}. 
We are grateful to 
the Institute for Nuclear Theory 
at the University of Washington, 
the SLAC Theory Group, 
the Summer Visitor's Program of the Fermilab Theory Group, 
and M.~B. Wise for hospitality 
during the completion of this work.

\end{document}